 \documentclass[final,5p,times,twocolumn,authoryear,sort&compress,square,comma]{elsarticle}

\usepackage{amssymb}
\usepackage{lipsum}
\usepackage{upgreek}
\usepackage{natbib}
\setcitestyle{numbers,square}

\journal{arXiv}

\begin{document}
\biboptions{numbers,sort&compress}
\begin{frontmatter}



\title{Light sheet and light field microscopy based on scanning Bessel beam illumination}


\author[first,second]{Chuhui Wang}
\author[second]{Jiaju Chen}
\author[second]{Cuiyi Peng}
\author[first,second]{Zhenglin Chen\corref{cor1}}
\ead{chenzlin1992@163.com}
\author[third]{Dongmei Yu\corref{cor1}}
\ead{yudongmei198011@sina.cn}
\author[first,second]{Peiwu Qin\corref{cor1}}
\cortext[cor1]{Corresponding author}
\ead{pwqin@sz.tsinghua.edu.cn}
\affiliation[first]{organization={Center of Precision Medicine and Healthcare, Tsinghua-Berkeley Shenzhen Institute},
            addressline={}, 
            city={Shenzhen,Guangdong Province},
            postcode={518055}, 
            state={},
            country={China}}
\affiliation[second]{organization={Institute of Biopharmaceutics and Health Engineering, Tsinghua Shenzhen International Graduate School},
            addressline={}, 
            city={Shenzhen,Guangdong Province},
            postcode={518055}, 
            state={},
            country={China}}
\affiliation[third]{organization={School of Mechanical, Electrical \& Information Engineering, Shandong University},
            addressline={}, 
            city={Weihai, Shandong Province},
            postcode={2642095}, 
            state={},
            country={China}}
\begin{abstract}
We developed a Bessel light sheet fluorescence microscopy (LSFM) system to enable high-speed, wide-field intra-vital imaging of zebrafish and other thick biological samples. This system uses air objectives for convenient mounting of large samples and incorporates an electrically tunable lens for automatic focusing during volumetric imaging. To enhance the precision of 3D imaging, the impact of the electrically tunable lens on system magnification is investigated and modified through designed experiments. Despite using Bessel beams with side lobes, we achieved satisfactory image quality through a straightforward background noise subtraction method, eliminating the need for further deconvolution. Our system provides zebrafish imaging at resolution comparable to commercial confocal microscopy but in just 1/40th of the time. We also introduced light field microscopy (LFM) to improve 3D in vivo imaging temporal resolution. Apart from the 28-fold speed enhancement, the comparison of LFM and LSFM results reveals a unique aspect of LFM imaging concerning image dynamic range, which has not been previously reported.
\end{abstract}



\end{frontmatter}




\section{Introduction}
\label{introduction}
Zebrafish have been extensively utilized as a model organism in scientific research, owing to their rapid, external development and remarkable optical transparency\cite{rf1,rf2,rf3,rf4}. The majority of zebrafish studies employ fluorescence imaging techniques, including confocal fluorescence microscopy\cite{rf5,rf6,dickensheets2018toward,liu2018mems}, multiphoton excitation microscopy\cite{rf7}, and light-sheet fluorescence microscopy (LSFM)\cite{rf8}. These three techniques all possess the ability of optical sectioning, which can help to improve the image quality and even achieve resolutions close to the optical diffraction limit. Both confocal microscopy and multiphoton microscopy employ point scanning. And confocal microscopy achieves optical section by incorporating two conjugate pinholes in the optical path. Although this can help to improve the resolution and signal-to-noise ratio (SNR) by eliminating the signals coming from out-of-focus areas, adjusting the conjugate pinholes during optical path construction can be cumbersome, and point scanning lowers the imaging speed. At the same time, because the sample is exposed to excitation light throughout the whole imaging process, significant photo-toxicity and photo-bleaching that are especially not suitable for in-vivo imaging are induced\cite{rf9}. Despite its inherent optical sectioning capability, multiphoton microscopy entails the use of costly lasers and specialized filters, rendering it less cost-effectively for numerous researchers\cite{rf10}. Furthermore, the exorbitant recurrent maintenance expenses render it financially prohibitive for numerous laboratories, simultaneously impeding the attainment of optimal equipment performance even with the equipment\cite{rf11}. Unlike the previous two methods, LSFM realizes optical sectioning by utilizing plane illumination\cite{rf12,rf13,rf14}. This not only can significantly improve the volume imaging speed, but also decrease the photo-damage and photo-bleaching drastically. The combination of these benefits and high spatial resolution renders light-sheet fluorescence microscopy (LSFM) widely used in biomedical investigations, particularly for intra-vital imaging of biological specimens\cite{rf15,rf16,rf17,rf18,rf19}.

In this paper, we developed a dynamic LSFM system for the fast 3D volumetric imaging of zebrafish. Due to the restricted field of view (FOV) of the conventional Gaussian light sheet, which can be generated by a cylindrical lens\cite{rf20}, non-diffracting beams are adopted to produce the excitation light sheet\cite{rf21,rf22}. Bessel and Airy beams are two representative non-diffracting beams\cite{rf23,rf24}. Because the generation of Airy beams necessitates the use of an expensive spatial light modulator (SLM)\cite{rf25}, Bessel beams, which can be produced by a more cost-effective axicon\cite{rf26}, are chosen for generating the excitation light sheet. Through this approach, a broad imaging FOV with uniform illumination, approximately 1.3 mm in size under 10X magnification detection objective, can be achieved. Compared with the 2-second per frame imaging speed demanded by commercial confocal microscopes and some other imaging methods\cite{rf27,rf28,rf29}, our LSFM is capable of enhancing the imaging speed by a factor of 40 with comparable lateral resolution. It should be emphasized that satisfactory image quality meeting application requirements can be achieved through simple background subtraction alone. There is no need for additional deconvolution. This represents an improvement for Bessel LSFM, as side-lobe effects necessitate deconvolution prior to obtaining satisfactory image quality\cite{rf30}. First-order defocus, resulting from the refractive index mismatch between the sample and medium, is another prevalent issue in volumetric imaging\cite{rf31}. To address this problem, an electrically tunable lens (ETL) is incorporated into the detection optical path to facilitate real-time focus adjustments during 3D imaging. In addition to the detailed method designed for adjusting the ETL to ensure auto-focus, the impact of ETL on system magnification is also investigated and calibrated to enhance the precision of the results. Apart from LSFM, we also incorporated light field microscopy (LFM) into the system to further enhance the time resolution. The light field information of the emission signals are recorded by LFM branch through a micro-lens array (MLA) as the conventional LFM\cite{rf32,rf33,rf34,rf35}. The basic theory of this branch is similar with selective volume illumination microscopy (SVIM)\cite{rf36}.

We apply this system in the imaging of zebrafish blood vessels and red blood cells circulation. LSFM results demonstrate that this cost-effective and simple post-processing system can provide a high-quality imaging platform for biomedical research involving zebrafish. It is important to note that the comparison of LFM and LSFM results reveals that while LFM's spatial resolution may not be on par with LSFM, LFM's signal decoupling function can help recover more detailed information in images with a high dynamic range where signal intensity is strong. To our knowledge, this is the first instance that the phenomenon has been identified. Furthermore, in our research, LFM, when utilizing traditional deconvolution 3D reconstruction methods, can improve 3D imaging speed by a factor of 28 compared to LSFM. In fact, our system serves as an effective imaging platform not only for zebrafish but also for organs and other thick biological samples.


\section{Methods}
\subsection{Optical setup}
\begin{figure}
	\centering 
	\includegraphics[width=0.4\textwidth, angle=0]{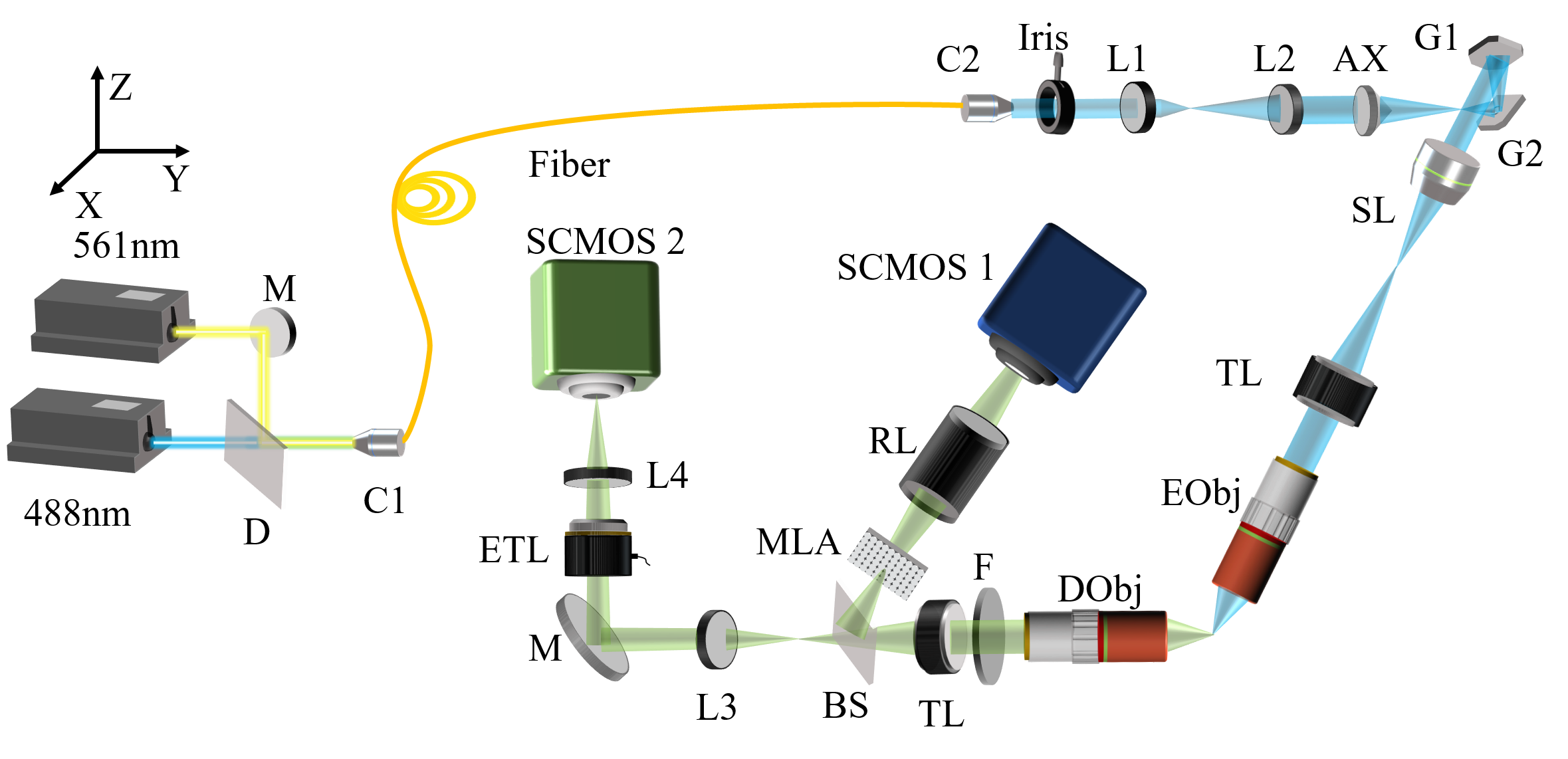}	
	\caption{Hybrid microscopy of light sheet and light field microscopy based on scanning Bessel beam. M: Mirror. D: Dichroic mirror. C1 and C2: Collimators. L1-L4: Achromatic doublets. AX: Axicon. G1-G2: Galvanometers. SL: Scan lens. TL: Tube lens. EObj and Dobj: Excitation and detection objective (Air). BS: Beam splitter. RL: Relay lens. F: Filter.} 
	\label{fig_1}%
\end{figure}

The schematic of the hybrid light sheet and light field microscopy based on scanning Bessel beam illumination is displayed in Fig. \ref{fig_1}. Two laser beams (488nm, 06-MLD, Cobolt, and 561nm, MGL-FN-561, Photon Tec Berlin) are coupled into a single-mode fiber and emit from a collimator. The figure takes the 488nm excitation light as an example to illustrate the propagation state of light waves after passing through various optical devices. The beam is expanded by a lens pair (L1 and L2) in a 4f configuration and then directed to an axicon (AX1210-A, Thorlabs) to generate a Bessel beam. Two-dimensional galvanometers (note that high speed scanners could be a better choice but more expensive\cite{liu2019mems}) G1 and G2 can scan the beam along the vertical direction (Z direction in the figure to form a light sheet in the focal plane of the detection objective) and the depth direction of the sample (Y direction in the figure). The excitation light then propagates to the sample through the scan lens, tube lens (TTL200MP, Thorlabs), and an air objective with 5x magnification (M Plan Apo NIR 5x/0.14, Mitutoyo). The emission signal is collected by the detective objective (M Plan Apo NIR 10x/0.26, Mitutoyo) and then propagates backward through the tube lens (TTL200-A, Thorlabs). A beam splitter splits the signal into two paths. One path is for light field imaging and the other is for light sheet imaging. In the LFM branch, the signal propagates through a micro-lens array (ML-S66-F1.0, Highlight Optics, China) and is transmitted to SCMOS1 (pco .edge 5.5, Germany) by an achromatic relay lens pair (MAP107575-A, 1:1 magnification, Thorlabs). In LSFM, L3 and L4 form a relay lens group and the ETL (EL-16-40-TC-VIS-20D-1-C, Optotune) is mounted between these two lenses. To overcome the influence of gravity on the liquid ETL, two mirrors (only one is drawn in Fig. \ref{fig_1} for convenience) are adopted to deflect the optical path in the vertical direction. SCMOS2 (ORCA Flash v2.0, Hamamatsu Photonics) is used to record the light sheet images. To mount the sample, a three-axis motorized translation stage (PT3/M-Z8, Thorlabs) is used, and a custom-designed sample mount based on a cuvette is assembled to the stage. 

To achieve three-dimensional scanning imaging, a custom LabVIEW control program was employed to interface the camera, translation stage, galvanometers, ETL, and a shutter positioned between D and C1 in Fig. \ref{fig_1} (not shown). During the LSFM imaging process, the Bessel beam is continuously scanned by G1 to form a Bessel light sheet in the front focal plane of the emission objective. The sample is moved by the translation stage in a predetermined fixed step, and the camera records images synchronously at each depth. To address defocusing caused by the mismatched refractive index and variable optical path during the scanning process, the current applied to the ETL also varies simultaneously with the stage movement. During the LFM imaging process, the sample remains stationary, while the Bessel beam is scanned not only in the vertical direction by G1 but also in the depth direction by G2, illuminating a specific volume of the sample near the focal plane of the detection objective. In general, the LFM imaging operation is similar to the concept of selective volume imaging microscopy.
\subsection{Sample preparing}
In this study, fluorescent beads of varying sizes and zebrafish serve as imaging samples. To image the beads, the sample is embedded in 0.5\% low-melting-point agarose. Prior to imaging the zebrafish, the sample is first anesthetized with tricaine (3-aminobenzoic acid ethyl ester, 0.2 mg/mL, Sigma Aldrich) for several minutes. Subsequently, the sample is drawn into a Fluorinated Ethylene Propylene (FEP) tube using a syringe. The FEP tube helps maintain the sample in a stretched and upright gesture. During imaging, the FEP tube containing the sample is secured by embedding it in 0.5\% low-melting-point agarose.
\subsection{The calibration of ETL}
During the 3D imaging process of LSFM, the change in the optical path between the sample on the focal plane of the detection objective and the camera causes defocus in the imaging results. An ETL is employed in the detection path to achieve auto-focus due to its rapid response capability (approximately 7 ms at 30 °C). The focal length of the ETL changes with the variation in applied current. In this context, two experiments are designed to determine the relationship between the ETL current and the stage position, as well as to explore the impact of the changing ETL focal length on the system magnification.
\subsubsection{Current-position curve measurement}
For 3D scanning, the sample is moved by an electrical position stage with a set fixed step continuously. Simultaneously, the focal length of the ETL should change with the position stage movement to ensure that all captured images are in focus. According to the product specification, the focal length of the ETL varies approximately linearly with the change in applied current. Therefore, there should be a relationship between the movement step of the positioning stage and the ETL current in the 3D scanning process. To describe this relationship in the LabVIEW-based control program, a module is designed to change the current applied to the ETL synchronously with the stage movement. Additionally, to ensure that this module achieves satisfactory autofocus effects in each scanning imaging experiment, it is necessary to measure the ETL current change for different samples or different areas of the same sample before each formal experiment. In the system testing stage, a homogeneous rhodamine solution, which can be excited by 488 nm and 561 nm lasers, is used as a sample. Detailed measurement steps are as follows:

(1)	Firstly, set the step size of the electrical position stage in the depth direction to a relatively larger value, such as 0.05 $\upmu$m (corresponding to a 5X excitation objective and a 10X reception objective), and set the current applied to the ETL to zero. Manually adjust the position of the receiving objective lens to find a plane with good focus of the sample, and record the current position display on the depth direction of the stage;

(2)	Move the stage in the depth direction according to the set step size, and the camera image will be out of focus relative to the original state. Adjust the current value on the ETL to make the image taken by the camera reach the focus plane closest to that in step (1), and record the position display of the stage at this time and the current value added to the ETL;

(3)	Repeat step (2) for eight to ten depth positions, noting down the coordinates of each depth position and the corresponding ETL current values that bring the image on the camera closest to the focusing state in step (1). Then measure the ETL current values at corresponding depths several times. For live sample imaging, to minimize light damage to the sample, the entire data sequence can be measured twice as quickly as possible.

(4)	Close the electric shutter to block the irradiation of the laser light source on the sample, average the current values of each depth position obtained in step (3) according to the number of measurements, and then plot the curve of current value against depth position. In actual experiments, it is observed that the curve exhibits a clear and good linear relationship. So the curve is fitted according to the linear relationship. The slope and intercept parameters of the fitted formula are input into the control program, enabling the focal length of the ETL to be controlled to vary synchronously with the positioning stage during the 3D scanning imaging process, thus achieving automatic focusing.

After completing the above steps and setting the corresponding parameters in the program, the defocusing of the image caused by the change in optical path can be effectively compensated when scanning the sample axially. It is important to note that in each experiment, the current-position curve of the ETL must be measured for different samples or even different areas of the same sample to achieve better auto-focusing results. Although preliminary measurements are included, auto-focusing compensation based on the ETL is more flexible and supports a larger axial scanning range compared to the compensation method using a liquid with a similar refractive index between the sample and the objective\cite{rf37}. 

As a uniform sample, the current-position curve of the rhodamine solution exhibits good linearity. Supplementary videos 1 and 2 demonstrate defocus and ETL-assisted auto-focus results, respectively. The bright line on the screen represents the profile of the Bessel beam. Please note that these two videos were captured under the condition that the optical axis of the detection path had not been aligned, causing the beam profile to move in the image during the auto-focus process. After realigning the optical axis of optical elements in the detection path, the light beam profile can be refocused with minimal movement on the screen. Supplement video 3 showcases the auto-focus of the rhodamine solution after optical axis alignment.

\begin{figure*}
	\centering 
	\includegraphics[width=0.8\textwidth, angle=0]{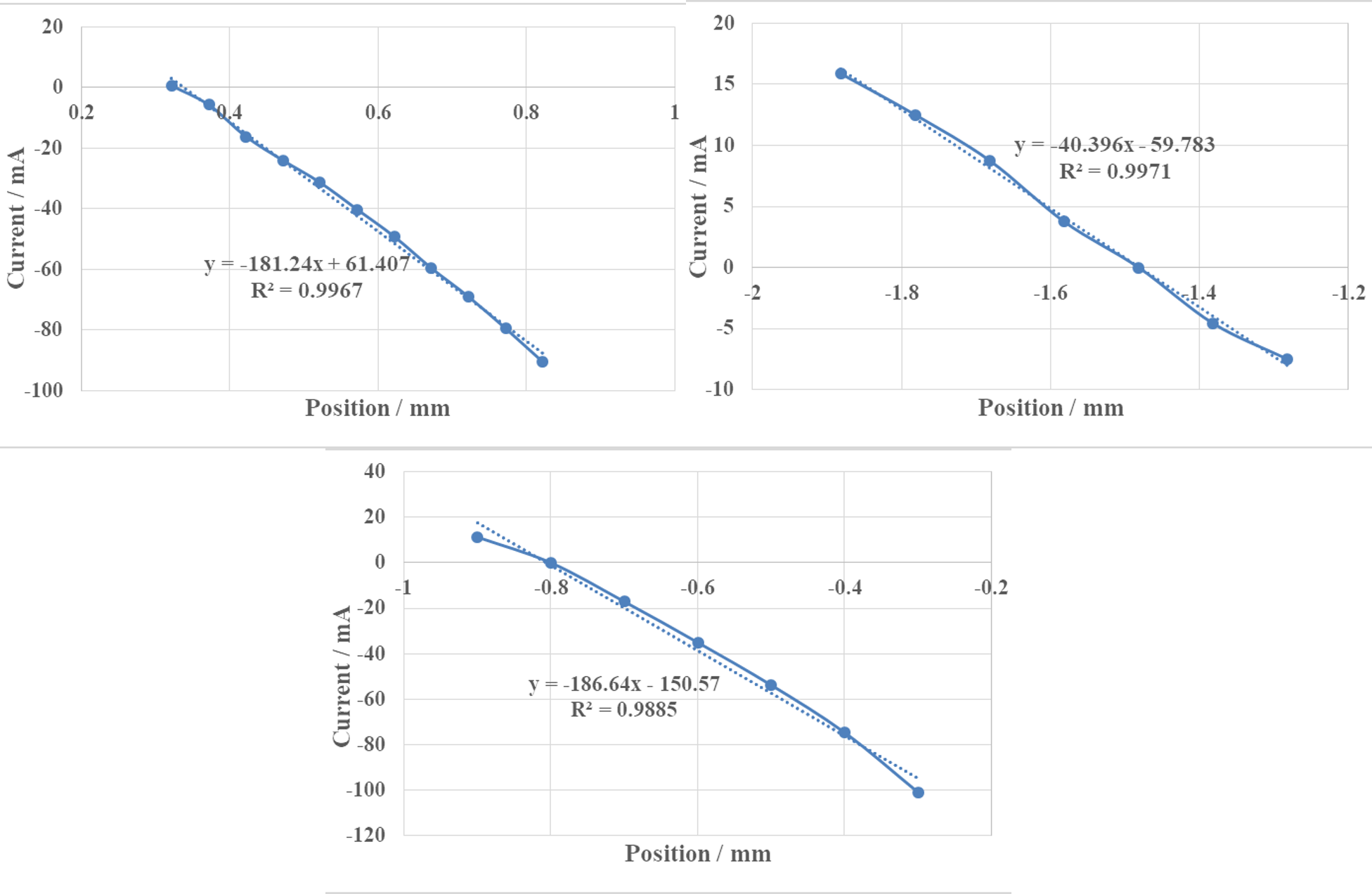}	
	\caption{The current-position curves of some experiments.} 
	\label{fig_2}%
\end{figure*}

In subsequent LSFM imaging experiments involving zebrafish in vivo, the measured current-position curves consistently exhibit good linearity. Fig. \ref{fig_2} illustrates the current position curves from some of these experiments. The fitting of these curves can be efficiently completed using readily available data processing software such as Origin or Excel.

\subsubsection{ Calibration of magnification}
Since the ETL is placed between a 4F system composed of two achromatic doublets with the original magnification ratio of 1:1, the change in its focal length will cause a change in the magnification ratio of the 4F system, thereby affecting the magnification of the image at different depths. Consequently, when performing volume fusion on images with varying depths, it is necessary to calibrate the magnification first. Here, we propose a calibration method using fluorescent beads with a relatively large particle size (diameter, 20 $\upmu$m) as follows:

(1)	Use fluorescent beads (20 $\upmu$m, Thermo Fisher) as samples. Apply a specific current to the ETL, then adjust the positioning stage to a plane where the sample is in focus. Record the image and label it as Image A;

(2)	Set the current of the ETL to zero and move the objective to achieve a focus state closest to that in step (1). Record the image of the current state and label it as Image B;

(3)	Change the current value applied to the ETL at a specific step and repeat steps (1) and (2) to obtain approximately 8-10 sets of images;

(4)	Process each set of images. First, select 2-5 beads with clear outlines in both Image A and Image B. Measure the diameter of each bead in both images. Calculate the ratio of the diameter measured in Image B to the corresponding diameter measured in Image A. Finally, average the ratio results of all selected beads in each group and label the value as R;

(5)	Draw a scatter plot with R as longitudinal and the corresponding current value as abscissa.

\begin{figure}
	\centering 
	\includegraphics[width=0.4\textwidth, angle=0]{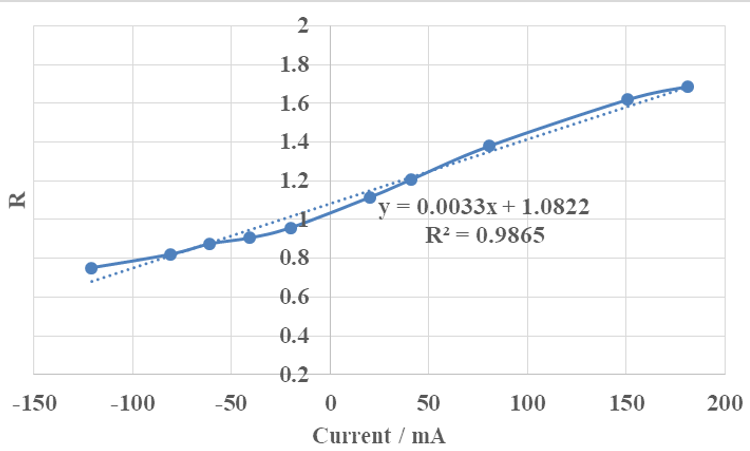}	
	\caption{Magnification and ETL current measurement curve.} 
	\label{fig_3}%
\end{figure}

By following the aforementioned five steps, a scatter plot illustrating the relationship between R and ETL current is generated. After fitting the curve equation, the R-value corresponding to any ETL current can be deduced. Fig. \ref{fig_3}  presents the experimental data and the fitting curve, which exhibits a linear trend.
It is crucial to emphasize that the curve is solely dependent on the current applied to the ETL. Unlike the current-position curve measurement process, which must be performed before each formal experiment, the R-current curve only needs to be measured once, and the fitting formula can be utilized for all experiments conducted with the same detection objective. By utilizing the R value and the refocused image A at one current, image B, which is ultimately used for volume synthesis in LSFM, can be obtained by resizing image A with magnification R.
\subsection{Volume synthesis of LSFM}
In the imaging of a relatively extensive biological specimen, such as the zebrafish, the sample is initially imaged along the depth axis (Y-axis in Fig. \ref{fig_1}) to acquire an image stack. Subsequently, the sample is translated in either the X or Z direction with a specific step size. This step must ensure an adequate overlap between the images before and after the sample's movement. The overlap facilitates the fusion of images in the same depth direction. Consequently, a series of image stacks with identical depth ranges and step sizes, but varying X or Y positions, can be obtained.

Utilizing the acquired stacks, we initially merge images at the same depth using the ImageJ stitching plugin\cite{rf38}. Subsequently, based on the initial position of the position stage, the moving step, and the current-position curve, the corresponding current for each depth can be calculated. Moreover, the corresponding R value for each depth can be obtained according to the R-current curve. In this context, the fused images at each depth are equivalent to image A in the magnification calibration step. By resizing these images using the R values, we obtain images with magnification that is minimally affected by ETL. The images are processed in batches using self-developed Matlab code. Prior to volume synthesis with these magnification-calibrated images, image size registration is required, as only images of the same size can be combined into a 3D volume. Considering that the resize function in Matlab is directly invoked during the magnification calibration step, an alternative image size registration method is employed in this case that does not affect the content of the images.

\begin{figure}
	\centering 
	\includegraphics[width=0.4\textwidth, angle=0]{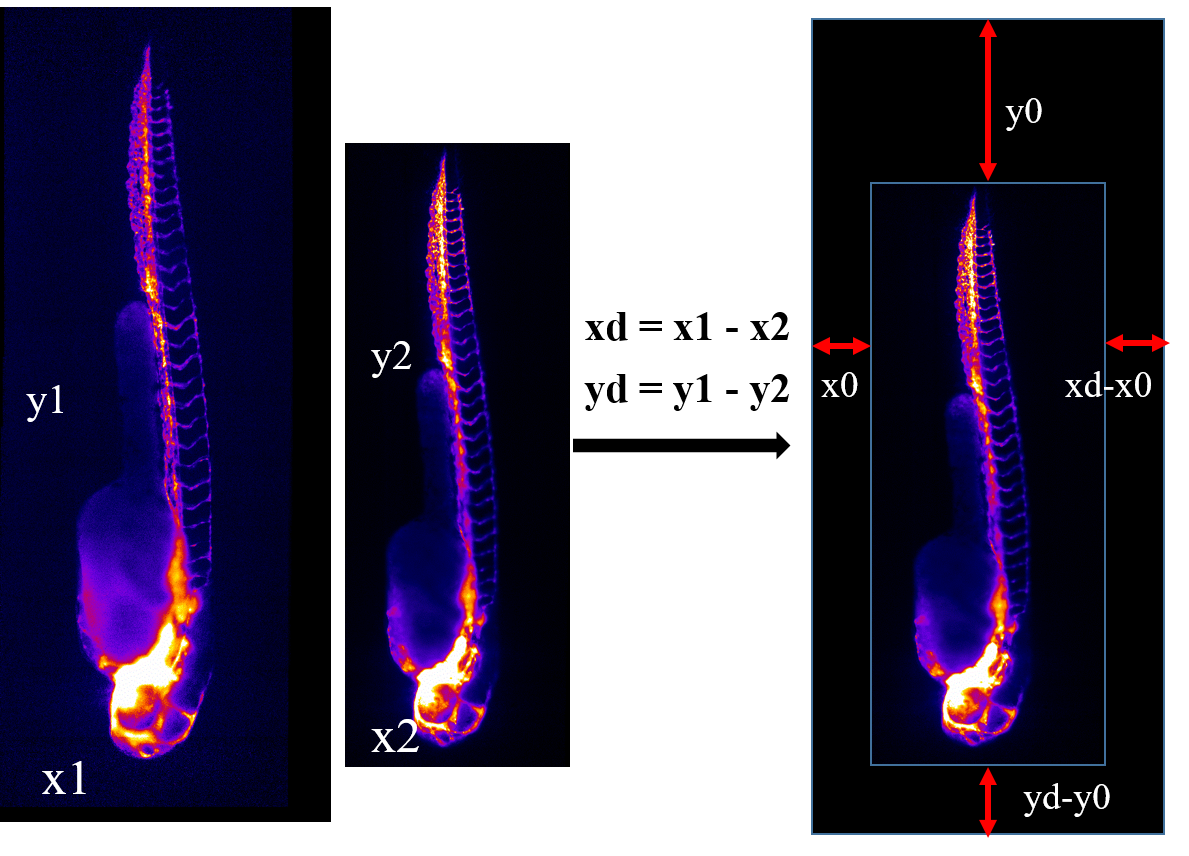}	
	\caption{Image size registration. x1, y1, x2, and y2 denote the sizes of the two images, while xd and yd represent the size differences. x0 and y0 are the mean values of the differences in the coordinates of the matched feature points} 
	\label{fig_4}%
\end{figure}

This is a three-step algorithm based on the classic SIFT (Scale-invariant feature transform)\cite{rf39}. Firstly, the image with the largest size is selected as a reference image, and for other images, feature points and their corresponding descriptions are extracted using the SIFT operator. Secondly, these descriptions and the K-nearest neighbor algorithm are used to determine the number of closest data points in space and categorize them into one class as a raw match\cite{rf40}. If the ratio of the closest point to the second closest point in the raw match exceeds a predetermined value (set as 0.6 in this case), the closest value is retained and considered a "good match." For these good matches, the coordinates are extracted, and the average difference in x and y coordinates is calculated as x0 and y0. Then is the last step that adjust the image size based on the calculation of the first two steps. Fig. \ref{fig_4} describes the image resizing operation.

\section{Results and discussion}
\begin{figure}
	\centering 
	\includegraphics[width=0.4\textwidth, angle=0]{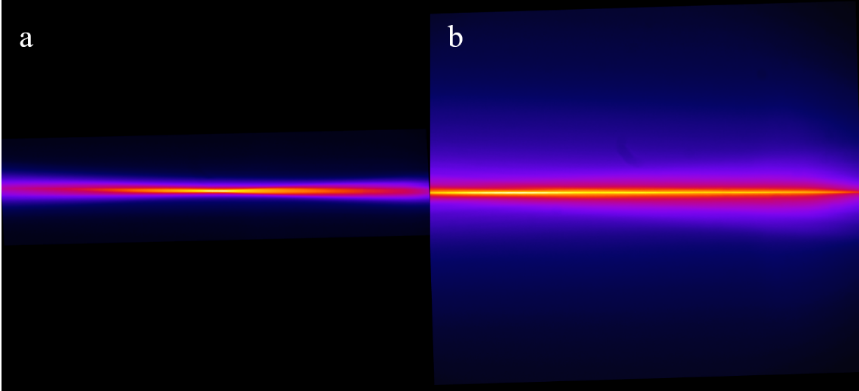}	
	\caption{The outlines of Gaussian light sheet and Bessel light sheet. a, Gaussian light sheet. b, Bessel light sheet.} 
	\label{fig_5}%
\end{figure}

Initially, we compare the Gaussian light sheet profile and the Bessel light sheet profile. The former is produced using a cylindrical lens (LJ1695RM, Thorlab), while the latter is generated with an axicon (AX1210-A, Thorlabs). The entire field of view (FOV) of the SCMOS camera (2048*2048 pixels) is utilized. The light sheet profiles are depicted in Fig. \ref{fig_5}.
Fig. \ref{fig_5}a illustrates a Gaussian light sheet profile with a full-width at half-maximum (FWHM) thickness of approximately 6.47 $\upmu$m (488 nm, wavelength-dependent) at the beam waist. Fig. \ref{fig_5}b displays a Bessel light sheet profile with an FWHM thickness of around 7.76 $\upmu$m (488 nm). Upon comparing these two images, it becomes apparent that although the background caused by sidelobes is more pronounced in the Bessel light sheet than in the Gaussian light sheet, the thickness of the Bessel light sheet is more uniform across the entire FOV. This confirms that non-diffracting beams are suitable for large FOV imaging. 

\begin{figure}
	\centering 
	\includegraphics[width=0.4\textwidth, angle=0]{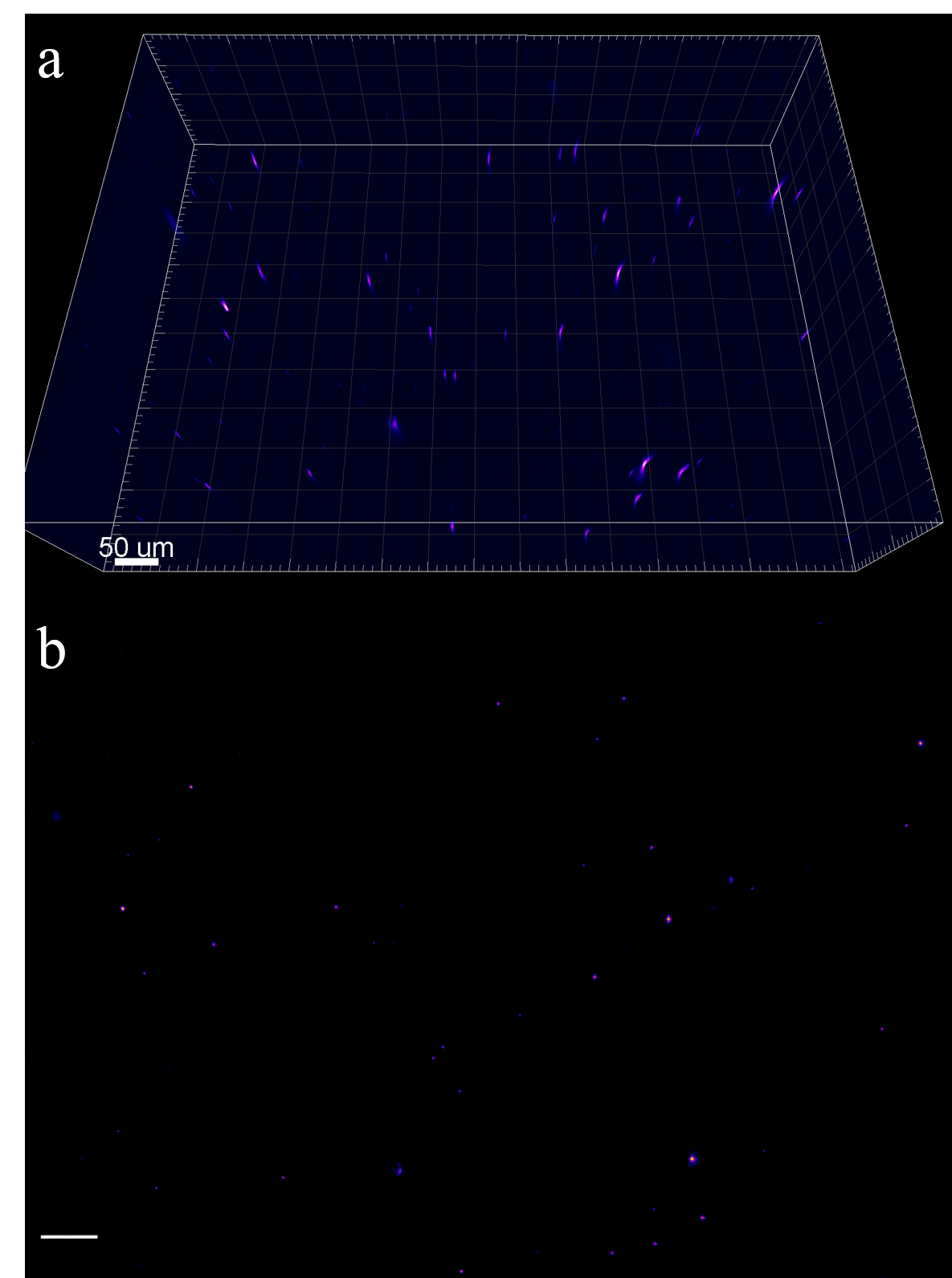}	
	\caption{PSF measurement of Bessel LSFM with 200 nm fluorescent beads. a, 3D volume reconstruction of beads. b, maximum intensity projection of beads image stack in the plane without depth direction. Scale bar, 50 um.} 
	\label{fig_6}%
\end{figure}

To validate the volumetric imaging capabilities of the scanning Bessel light sheet illumination LSFM and LFM, 3D imaging experiments were conducted using fluorescent beads and zebrafish embryos. Fig. \ref{fig_6} presents the point spread function (PSF) measurement results of LSFM with 200 nm fluorescent beads. The resolutions of the images were measured without deconvolution. The lateral resolution is approximately 2.06 $\upmu$m, and the axial resolution is approximately 12.76 $\upmu$m. The lateral resolution is sufficient for the detection of blood cells.
The zebrafish lines Tg (fli1a: EGFP) and Tg (gata1a: DsRed) serve as two types of samples. The former has become a crucial tool for investigating zebrafish vascular development and can be excited by a 488 nm laser. The latter enables continuous tracking and observation of red blood cells under the excitation light of 561 nm. Initially, we examine the vascular structure of zebrafish using 488 nm excitation light. Both LSFM volume imaging and LFM imaging are performed. 

\begin{figure*}
	\centering 
	\includegraphics[width=0.6\textwidth, angle=0]{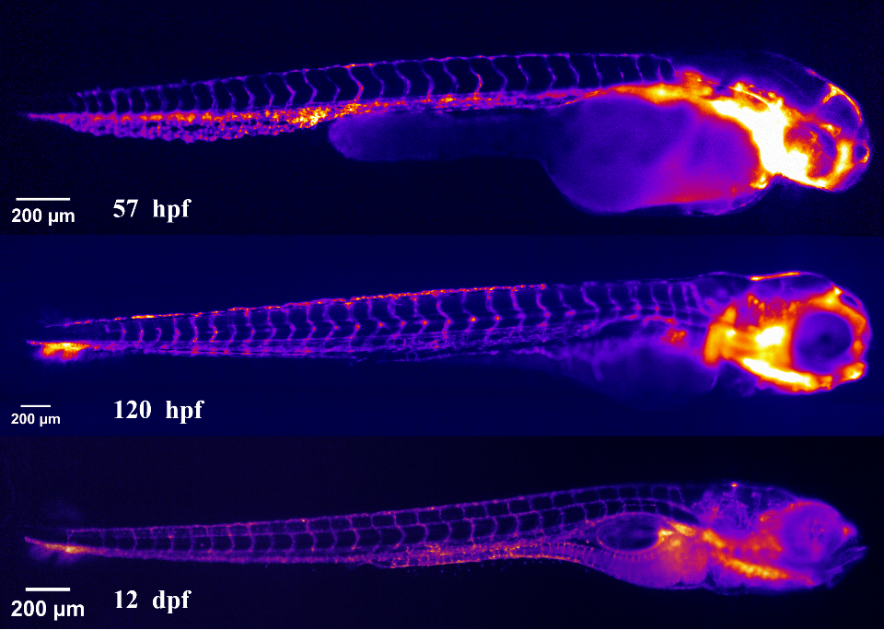}	
	\caption{ Single-depth image of zebrafish blood vessel structures at different stages. hpf, hours post-fertilization; dpf, days post-fertilization.} 
	\label{fig_7}%
\end{figure*}

During the LFM imaging process, the Bessel beam is rapidly scanned not only in the XZ plane but also in the Y direction, enabling the illumination of a sample volume and the recording of signals within this volume by the camera in a single snapshot. In the LSFM imaging process, the Bessel beam is only fast-scanned in the XY plane to create a light sheet. Fig. \ref{fig_7} displays the vascular structure of zebrafish at various depths, imaged using LSFM. It is important to note that all LSFM images of zebrafish in this study are post-processed solely by background subtraction to enhance contrast and signal-to-noise ratio (SNR). In 2D imaging of zebrafish at a single depth, the background is recorded by moving the sample mount to a position where the sample is not illuminated by the excitation light and capturing an image of the media. In 3D imaging of zebrafish, the background stack is recorded using the same imaging operation as the sample, and background subtraction is performed by subtracting the background at the corresponding depth from the original image. A self-developed Matlab code is employed for batch processing of background subtraction. No further deconvolution of LSFM is conducted.

\begin{figure*}
	\centering 
	\includegraphics[width=0.6\textwidth, angle=0]{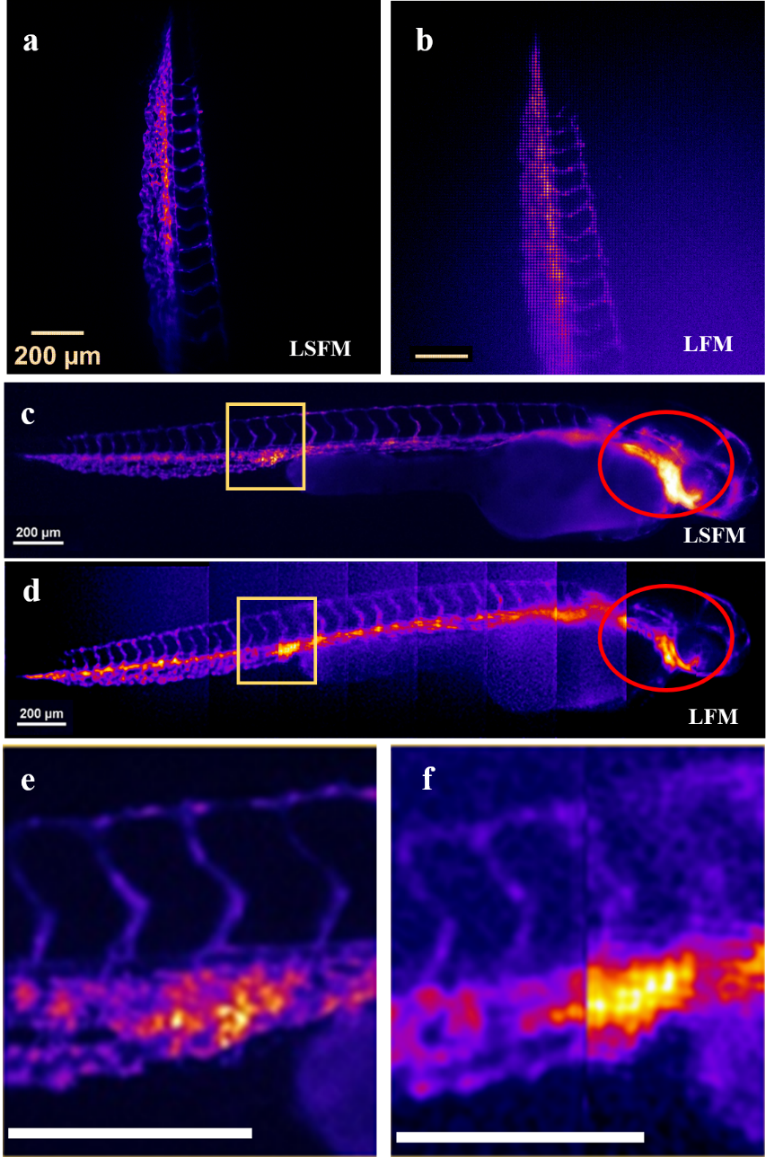}	
	\caption{Comparison of imaging results of LSFM and LFM. a, one single LSFM image of a zebrafish tail. b, the original LFM image of the zebrafish tail. c,d, vascular structure imaging results of LSFM and LFM reconstruction. e,f, magnification of yellow rectangular areas in c and d to show the resolution difference. Note, the vertical line in f is caused by imperfect image stitching. Scalebar, 200 um.} 
	\label{fig_8}%
\end{figure*}

\begin{figure*}
	\centering 
	\includegraphics[width=0.8\textwidth, angle=0]{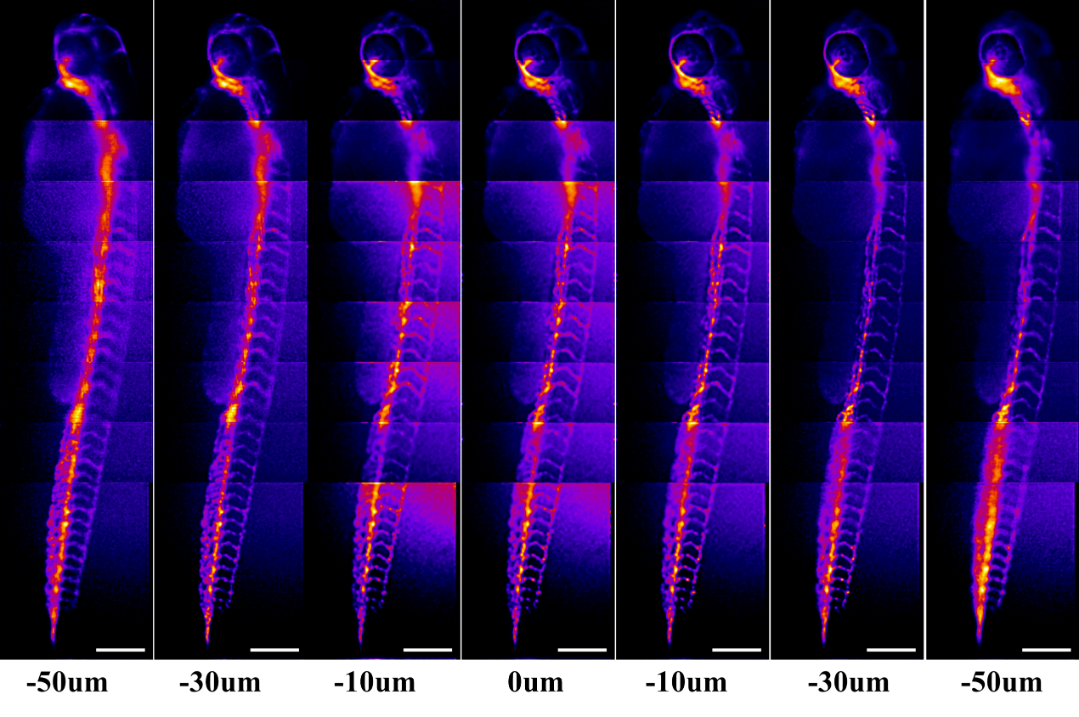}	
	\caption{ The 3D reconstruction results of LFM in different depths.} 
	\label{fig_9}%
\end{figure*}

Fig. \ref{fig_8}a represents a single frame of an LSFM image, while Fig. \ref{fig_8}b shows an original LFM image. Due to the sample signal being split with a beam splitter (1:1 ratio), this increases the exposure time for both LSFM and LFM to some extent. To ensure adequate SNR, the exposure time for each LSFM and LFM image is set to 50 ms. Fig. 8a captures the tail part of the sample in a single shot. After moving the sample along the Z direction and recording the signals of all samples at this depth, the 2D vascular structure of zebrafish in Fig. \ref{fig_8}c can be obtained through image fusion. Fig. 8b is the original light field image of the sample, and Fig. 8d and Fig. \ref{fig_9} are reconstruction results of LFM at different depths. Fig. \ref{fig_8}d is the layer closest to the LFSM result in Fig. \ref{fig_8}c. There are some noise shadows in Fig. \ref{fig_8}d and Fig. \ref{fig_9}, originating from background noise and amplified during the RL-deconvolution. Despite the presence of noise in LFM, the blood vessels are still discernible enough to identify the overall structure. Comparing Fig. \ref{fig_8}c and Fig. \ref{fig_8}d, it is evident that the reconstruction resolution of LFM is lower than that of LFSM. For instance, the rectangular areas in Fig. \ref{fig_8}c and Fig. \ref{fig_8}d, magnified in Fig. \ref{fig_8}e and Fig. \ref{fig_8}f, display the resolution difference. The lower reconstruction resolution is attributed to the LFM imaging method, where the optical signal is sampled by MLA below the Shannon-Nyquist limit, leading to high-frequency feature aliasing and blurring of details in the reconstruction results. Comparing LSFM and LFM imaging results, an interesting phenomenon is observed. In the red circle area in Fig. \ref{fig_8}c and Fig. \ref{fig_8}d, where blood vessels are densely distributed, LSFM images are prone to overexposure due to stronger signal strength compared to other areas. In contrast, LFM's dispersed and relatively weaker signals prevent overexposure of these areas, allowing for the display of details in LFM results, albeit not as clear due to LFM reconstruction resolution limitations. This imaging potential of LFM can be further enhanced by employing MLA with a larger lens size or modifying the LFM reconstruction method to improve reconstruction quality. This phenomenon may stem from LFM's imaging mode, as light emitted from the same point on the object in different directions is received by different camera pixels and used for 3D reconstruction. Consequently, the signal on each slice of the reconstructed 3D sample is weaker than that of LSFM, which receives the signal in a wide-field mode. This helps reduce the dynamic range of the images and prevents signals from being so strong that most details cannot be distinguished.

\begin{figure*}
	\centering 
	\includegraphics[width=0.6\textwidth, angle=0]{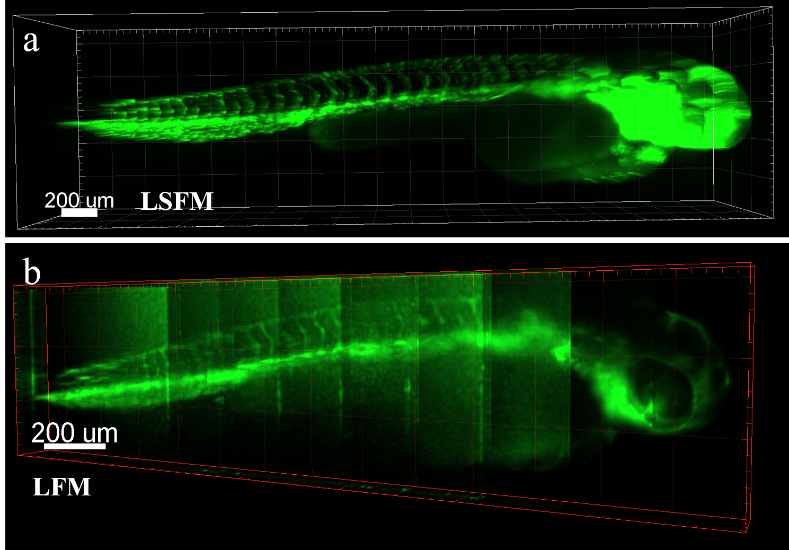}	
	\caption{ 3D reconstruction of the vascular structure of zebrafish. a, LSFM imaging. Depth range, about 0.6 mm. b, LFM reconstruction results. Depth range, about 0.1 mm.} 
	\label{fig_10}%
\end{figure*}

Apart from the 2D plane of the sample, 3D volume imaging are also constructed. Fig. \ref{fig_10} showss 3D reconstruction results of zebrafish blood vessel structures in both LSFM and LFM. In LSFM, the imaging range in the depth direction is 600 $\upmu$m, and 150 images were taken with 4 $\upmu$m steps. To reconstruct the volume and save image size registration time, a total of 50 images were selected with a slice spacing of 12 $\upmu$m. Fig. \ref{fig_8}a presents the reconstruction results of LSFM, and supplementary video 4 demonstrates the heart beating of zebrafish recorded by LSFM. In LFM, the volume is reconstructed using the artifact-free Richardson-Lucy deconvolution method\cite{rf41}. Due to the fact that each small lens of the micro-lens array only covers approximately 10 pixels of the SCMOS camera, the perspective information collected by LFM is limited to some extent. As a result, the reconstruction range in the depth direction of LFM is only about 0.1 mm. Beyond this range, numerous block artifacts and blurs occur, leading to poor image quality. Fig. \ref{fig_8}b displays the LFM 3D reconstruction results. Although local noise from Richardson-Lucy deconvolution degrades the quality, LFM can effectively improve 3D imaging speed by 25 times compared to LSFM (0.1 mm range, 4 $\upmu$m steps).

\begin{figure*}
	\centering 
	\includegraphics[width=0.6\textwidth, angle=0]{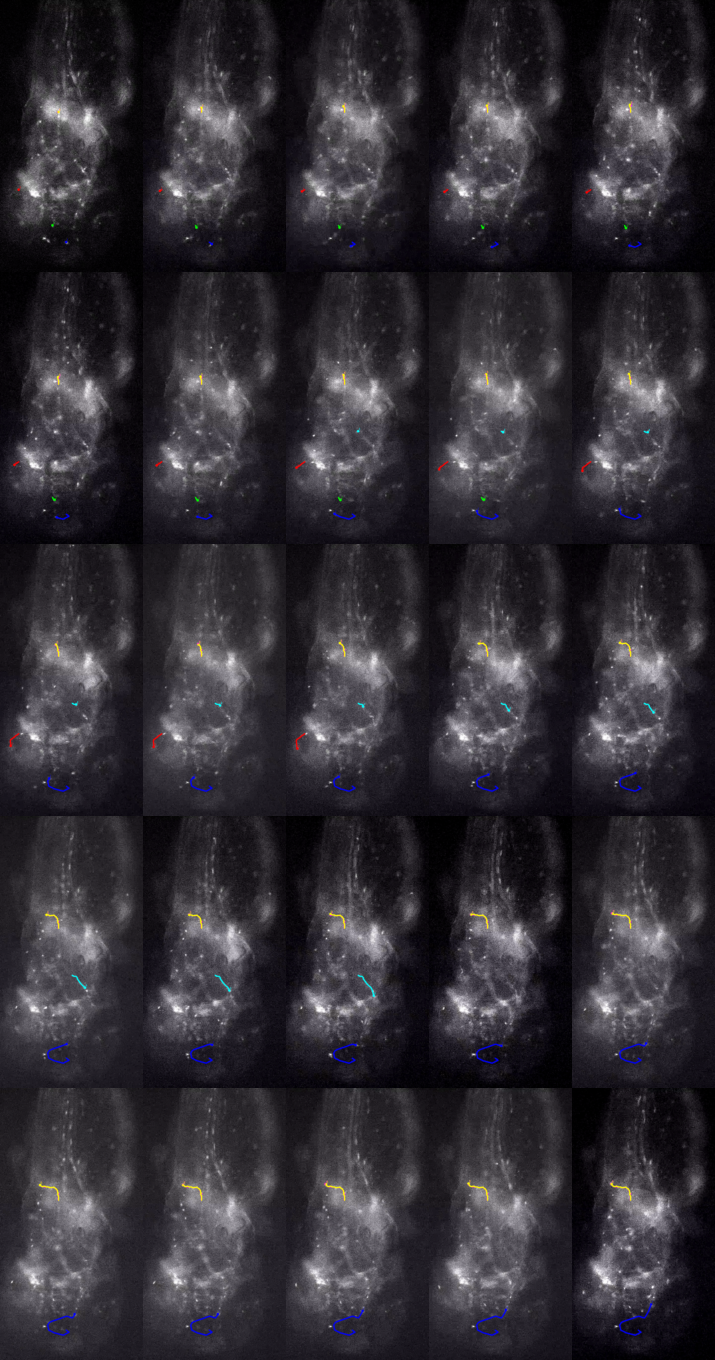}	
	\caption{ LSFM imaging zebrafish red cell flow process. Curves with different colors represent the trajectories of different red cells.} 
	\label{fig_11}%
\end{figure*}

Transgenic zebrafish labeled with red blood cells (Tg (gata1a: DsRed)) were imaged using 561 nm excitation light to observe zebrafish blood flow. Supplementary video 5 demonstrates the blood flow near the zebrafish heart recorded with LSFM. Fig. \ref{fig_11} displays the trajectories of several red blood cells in a time series of images, with different cells represented by different colors. This confirms that our LSFM system is suitable for hemodynamic imaging experiments involving zebrafish.

\section{Summarization}
In summary, a scanning Bessel light sheet illumination LSFM and LFM combination imaging system has been developed. This system is primarily designed for imaging biomedical samples of relatively large size, such as zebrafish embryos, organoids, and nematodes. Air objectives with long working distances are employed to facilitate sample mounting. In LSFM, a 3D volume imaging pipeline was developed, including ETL auto-focus registration, sample preparation and fixation, image magnification correction, and volume synthesis. This enabled the mapping of the zebrafish vascular structure without time-consuming and costly deconvolution, providing both a complete two-dimensional plane structure map at a single depth and a volume imaging map within a specific range (~0.73*3.16*0.6mm) at cellular resolution. For LFM, the volume vascular structure within a certain range (~0.73*3.16*0.11mm) was reconstructed using the classical deconvolution method. Although the LFM reconstruction resolution is lower than expected, it was found that in areas with strong fluorescence signals, LFM results exhibit higher contrast than LSFM and offer nearly 28-fold faster imaging speeds.

Compared to LSFM, which generates a light sheet using a SLM, our system is cost-effective and easily adjustable. In comparison to confocal microscopy, which can achieve similar image quality, our LSFM can enhance imaging speed by 40 times, and LFM can improve imaging speed by several hundred times. We hold the belief that this system can function as an exceptional detection platform for biomedical research, particularly when employing zebrafish as a model organism\cite{rf42}. Further optimizations and explorations of the system, such as enhancing the image quality of LSFM and LFM by integrating deep learning methods\cite{rf43,rf44,rf45,rf46,rf47}, are currently underway.

\section{Acknowledgements}
We would like to thank Professor Dong Liu, Dongmei Su and Dandong Yuan (Southern University of Science and Technology) for supplying zebrafish samples used in the research. 
\section{Ethics Statement}
The animal study was reviewed and approved by Ethical Committee of Tsinghua Shenzhen International Graduate School.
\section{Declaration of Competing Interest}
The authors declare that they have no competing financial interests or personal relationships that could have appeared to influence the work reported in this paper.
\section{Funding}
We thank the support from the National Natural Science Foundation of China 31970752; Science, Technology, Innovation Commission of Shenzhen Municipality JCYJ20190809180003689, JSGG20200225150707332, JCYJ20220530143014032, ZDSYS20200820165400003, WDZC20200820173710001, WDZC20200821150704001, JSGG20191129110812708, KCXFZ20211020163813019; Shenzhen Bay Laboratory Open Funding, SZBL2020090501004; Department of Chemical Engineering-iBHE special cooperation joint fund project, DCE-iBHE-2022-3; Tsinghua Shenzhen International Graduate School Cross-disciplinary Research and Innovation Fund Research Plan, JC2022009; and Bureau of Planning, Land and Resources of Shenzhen Municipality (2022) 207.

\bibliographystyle{elsarticle-num}
\bibliography{ref}






\end{document}